# Platelet-zone in an age-hardening Mg-Zn-Gd alloy

T. Koizumi[a], M. Egami[a,b], K. Yamashita[a], E. Abe [a,b*]

[a]*Department of Materials Science & Engineering, University of Tokyo, Tokyo 113-8656, Japan*
[b]*Research Center for Structural Materials, National Institute for Materials Science 305-0047, Japan*
*Corresponding author : abe@material.t.u-tokyo.ac.jp

**Abstract**

The structure of a unique platelet zone with a three close-packed layer thickness, which occurred in a Mg-1at.%Zn-2at.%Gd alloy annealed at low temperatures (<~500K), has been determined based on scanning transmission electron microscopy and first principles calculations. The platelet zone is constructed by the definite Gd networks that cause significant atomic displacements in the adjacent close-packed layer, in which the atomic sites are preferentially occupied by Zn atoms. Local Zn concentrations at the reconstructed-layer have been successfully tuned according to a similar manner of Vegard's law by referring the local Gd-Gd interatomic distance along the *c*-axis. Therefore, the present platelet zone is found to be not a simple ordered hexagonal-close-packed Mg-Gd as previously reported, but a reconstructed ternary Mg-Zn-Gd structure.

*Keywords:* Magnesium alloys; Fine precipitations; Atomic structure; Electron microscopy; First principles calculations

## 1. Introduction

Recently, Mg-TM-RE (TM: transition metals, RE: rare earth elements) ternary alloys have been of focused interest as one of the promising lightweight structural materials for next generations. Particularly, several $Mg_{97}TM_1RE_2$ (at.%) alloys revealed excellent mechanical properties by having a unique long period stacking/order (LPSO) phase [1-10] as a strengthening phase through kink-band formation [11-13]. For significant age-hardening Mg-Zn-Gd alloys [14, 15], the LPSO as well as a series of the *β* phases indeed precipitate from the supersaturated solid-solution (SSSS) Mg matrix. Besides, it was also found that unique solute-enriched platelet zones (Guinier-Preston (GP)-like zone) densely form on the (0001) planes of the hexagonal-close-packed (hcp) matrix structure [16], which become prominent when the alloys are aged at relatively low temperatures (<~500K). Note that a rich occurrence of these various phases/zones (see Fig. 1a) are provided by Zn additions, since the LPSO phase and the plate zones do not occur for the Mg-Gd binary alloys.

The GP-like platelet zones in the Mg-Zn-Gd alloys are found to contain essentially both the Zn and Gd atoms and have been structurally classified into two types; i.e., solute-enrichment occurs at the (i) original AB stacking planes of the hcp structure [15, 16] and (ii) intrinsic-II type ABCA stacking fault planes [14, 15]. As for the latter case (ii), the solute-enriched ABCA stacking is basically identical with the structural unit of the LPSO structures [5-7], and it is referred to as stacking-fault zones (SFZ; though being referred to as simply "stacking fault" in Ref. 14, it is obviously a growing "zone" during isothermal annealing). On the other hand, for the case (i), it is shown that the GP-like zones are not due to random clustering of solute atoms but constructed by the definite ordered arrangement of Gd atoms [16], which are placed at the second-nearest atomic sites in the hcp host structure. Hence it was referred to as ordered GP zones [16], but their detailed structure including Zn atom positions have not been clarified yet.

In the present work, we investigate the structure of the so-called ordered GP zones in the Mg-Zn-Gd alloys, based on atomic-resolution scanning transmission electron microscopy (STEM) and first principles calculations. As will be described in the following, the present platelet zone has been determined to be a well-ordered ternary Mg-Zn-Gd structure with a three close-packed layer thickness, in which significant atomic displacements take place at the middle close-packed layers. Therefore, the present zone is not a simple ordered hcp structure but reasonably viewed as a reconstructed crystal of a unit-cell thick. In this regard, we will term the present zone as the platelet-crystalline (PC) zone, in order to distinguish from the well-known GP zones that basically refer random clustering (or atmosphere) of solute atoms at the particular host lattice planes (e.g., {100} planes of a face-center-cubic lattice in Al alloys).

## 2. Experiment

An alloy with the nominal composition of $Mg_{97}Zn_1Gd_2$ (at.%) was prepared by high-frequency induction melting of Mg, Zn and Gd in a carbon crucible. The alloy ingots were annealed at 473K for 200h. Thin foils for STEM observations were prepared by a standard argon ion milling technique. STEM observations were performed by an aberration-corrected 200kV microscope (JEM-ARM200F) equipped with a cold filed-emission gun. The annular detectors were set to collect electrons scattered at angles between 90 and 370 mrad for high-angle annular dark-field (HAADF) imaging, and 11 and 22 mrad for annular bright-field (ABF) imaging [17, 18]. Image simulations were carried out using WinHREM software. First principles calculations were performed using the Vienna Ab initio Simulation Package (VASP) within the framework of density functional theory (DFT), based on generalized gradient approximation (GGA) and ultrasoft scalar relativistic pseudo-potential.

## 3. Results and discussion

Figures 1a and b show the low-magnification HAADF-STEM image and the corresponding electron diffraction pattern obtained from the present $Mg_{97}Zn_1Gd_2$ alloy. Since the HAADF imaging provides the significant atomic-number (Z) dependent contrast, in the image the bright area directly represent the Gd and/or Zn enriched zones (Z: Mg = 12, Zn = 30 and Gd = 64). It is clearly seen that the microstructure consist of a variety of fine precipitates and indeed provides the almost maximum hardness during aging [14]. Among the precipitates in Fig. 1a, some diffuse contrast well developed along the *c*-axis represent the *β'*-precipitates, which give rise the weak spots along the 1/2 positions to the $(11\bar{2}0)_{hcp}$ in the diffraction pattern of Fig. 1b. On the other hand, there definitely exist two types for the platelet zones revealing sharp line-contrast; stacking-fault (SF) and platelet-crystalline (PC) zones, which are clearly seen in the atomic-resolution HAADF images shown in Fig. 1c and d. It is immediately noticed in Fig. 1d that the stacking sequence in the PC zone is not simply of hcp/fcc type. The PC zones reveal the suprlattice streaks at 1/3 and 2/3 positions with respect to the $(11\bar{2}0)_{hcp}$ reflections, as indicated by arrows in Fig. 1b, while the SF zones simply give rise streaks along the $c^*$-directions for the fundamental reflections. The present work focuses on the detailed atomic structure of the PC zones, which are identical with those referred to as the ordered GP zones [16].

For STEM observations, we employ an effective combined use of HAADF/ABF imaging; the former highlights the heavy atom positions, while the latter provides a high-sensitive imaging to detect the light atom positions, providing a comprehensive method for real-space determination of the multicomponent structures [19]. Figs. 2 a-d show the atomic-resolution HAADF/ABF images of the PC zones, identifying clearly that they are distinctly composed of three close-packed atomic layers. For the HAADF images, the brightest dots represent the atomic columns containing Gd atoms, the distribution of which is basically identical with those observed in the previous HAADF image [16]. On this basis, the ordered Gd arrangements were directly deduced by simply placing Gd atoms on the original hcp sites. However we note in particular that, in the present ultrahigh-resolution HAADF/ABF images, some atomic sites within the zones are not coincident with those of the host hcp structure. In addition, the distance between the brightest atom columns (i.e., Gd-Gd interatomic distance) along the $c$-axis is estimated to be about 3.8Å by the HAADF images (see the intensity profiles attached to Figs. 1 a and b), which is remarkably shorter than that expected from the hcp-Mg structure ($c_{Mg} \sim 5.2$ Å). These features suggest that the PC zone structure is not simply an ordered hcp structure [16], and significant atomic relaxations/reconstructions take place within the PC zone.

During attempting to tune the present PC zone structure with the aid of first principles calculations, we find that significant reconstructions immediately take place for the initial Mg-Gd configurations. The initial input is a hcp-based structure by placing Gd at the second-nearest positions in the relevant close-packed planes (Fig. 3, left), which turns out to be significantly reconstructed after energetic optimizations, as shown right in Fig. 3. Among the three close-packed layers, the remarkable atomic displacements occur at the middle layer, as marked by semitransparent-orange in Fig. 3, in which the Mg atoms form large six-membered ring to make space for the Gd atoms at the adjacent layers. In addition, the Gd atoms shift inward along the $c$-axis to form the puckering layers, and consequently the interatomic Gd-Gd distance ($d_{Gd-Gd}$) becomes shorter (~4.1Å). However, the $d_{Gd-Gd}$ value in Fig. 3 seems to be not yet optimized by comparing with the observed $d_{Gd-Gd}$ ~3.8Å. Looking carefully at the layer-by-layer HAADF intensity profile shown in Figs. 1 a and b, the middle-layer indeed reveals stronger intensity than that of the Mg matrix, suggesting that the Zn atoms are likely to occupy the middle layer. Remembering that the atomic radius of Zn, $r_{Zn} \sim 1.34$ Å, is smaller that that of the other atoms ($r_{Mg} \sim 1.60$ Å, $r_{Gd} \sim 1.80$ Å), Zn atoms are favorable to replace the Mg-site at the reconstructed middle-layer where the Mg-Mg intervals are forced into a shorter distance ~ 2.83Å (Fig. 3). Zn occupations at the middle-layer would then lead to shrinking the $d_{Gd-Gd}$ along the $c$-axis. Along this concept, we further tune the PC zone structure by changing the Zn concentrations at the middle layer, and the results are plotted with respect to the $d_{Gd-Gd}$ along the $c$-axis, as shown in Fig. 4. It turns out that there is a good linear correlation between them, and this can be used for evaluation of the Zn-concentration along with a similar manner of a Vegrad's law; the Zn concentrations at the middle layer is estimated as being about ~60% to fit the observed $d_{Gd-Gd}$ value of 3.8Å (Fig. 1). It should be noted here that, although the present estimation is based on ground state configurations

(i.e., atomic configurations at 0K), temperature effects may be reasonably negligible because of a small value of thermal lattice expansions of a Mg crystal. That is, a linear coefficient of thermal expansion of the Mg alloys is generally in a order of $\sim 10^{-5}$/K (e.g., $\sim 2.7 \times 10^{-5}$/K for AZ-Mg alloys), which results in at most ~1% increases of the lattice parameter (a order of ~0.01Å) even at the present annealing temperature 473K.

A validity of the present PC zone model is then confirmed based on HAADF/ABF atomic-image simulations, by which all the atom positions of Mg, Zn and Gd in the present PC zone can be almost unambiguously identified [19]. Figure 5 shows a series of the experimental data, simulation results and the corresponding projections of the model structure. As seen in Figs. 5 a - j, all the observed characteristics have been reproduced fairly well both for the HAADF/ABF images; 1) extra atomic sites generated at the middle layer of the zone, 2) intensity distributions at each of the atomic columns, and 3) a puckering feature for the Gd-containing close-packed layers.

As described above, the present PC zones appear to be a definite Mg-Zn-Gd crystalline order within the three close-packed layers. Remembering that the isolated SF zones (Fig. 1d) are able to construct the LPSO structure [5-7], the present PC zone could be a potential unit to form a long-range order structure by stacking them along the *c*-axis, perhaps occurring in some of the ternary Mg-TM-RE alloys. In fact, we find that the long-period superlattice (LPSL) Mg-Zn-Yb structure recently identified [20] has the almost identical local RE configurations/networks as those in the present reconstructed-layer (Fig. 3), and further that the second LPSL phase provides a direct example of the long-range ordered crystalline state of the present PC zones [21]. Investigations of a novel series of LPSL-Mg-TM-RE phases are now in progress, and the results will soon be described elsewhere.

## 4. Conclusion

In summary, we have determined unambiguously the structure of the PC zone with a three close-packed layer thickness, which occurred in a Mg-1at.%Zn-2at.%Gd alloy annealed at 473K. The zone structure is well characterized by a definite Gd network and the reconstructed middle layer, where Zn atoms preferentially occupy the site by approximately 60%. It is anticipated that the present PC zone could be a possible building-unit to construct a novel series of long-period superlatice structures in Mg-TM-RE alloys.

**Acknowledgment**

This study is supported by the Grants-in-Aid Scientific Research on Innovative Areas "Synchro-LPSO Structures" (No. 23109007) and "Nanotechnology Platform" (No.12024046) from MEXT/JSPS, Japan.

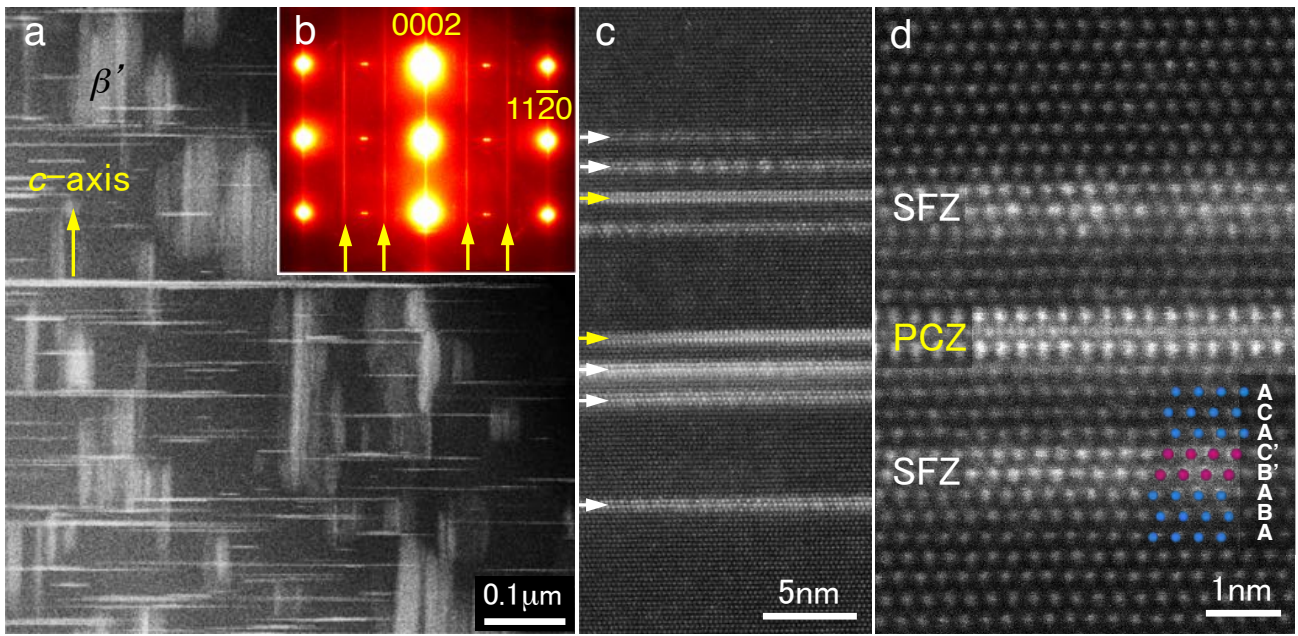

**Figure 1.** (a) Low-magnification HAADF-STEM image and (b) the corresponding electron diffraction pattern taken along the $[1\bar{1}00]_{hcp}$ direction, obtained from the $Mg_{97}Zn_1Gd_2$ alloy aged at 473K for 200h. (c) High-magnification and (d) atomic-resolution HAADF images taken along the $[11\bar{2}0]_{hcp}$ direction, showing distributions of the platelet zones which are classified into two types; stacking-fault zone (SFZ) and platelet-crystalline zone (PCZ) as indicated by white and yellow arrows in (c). SFZ is a structural-unit of the LPSO structure (ref. 5-7), and its atomic model is inserted in (d) where the Zn/Y atoms are significantly enriched at the close-packed layers B'C'.

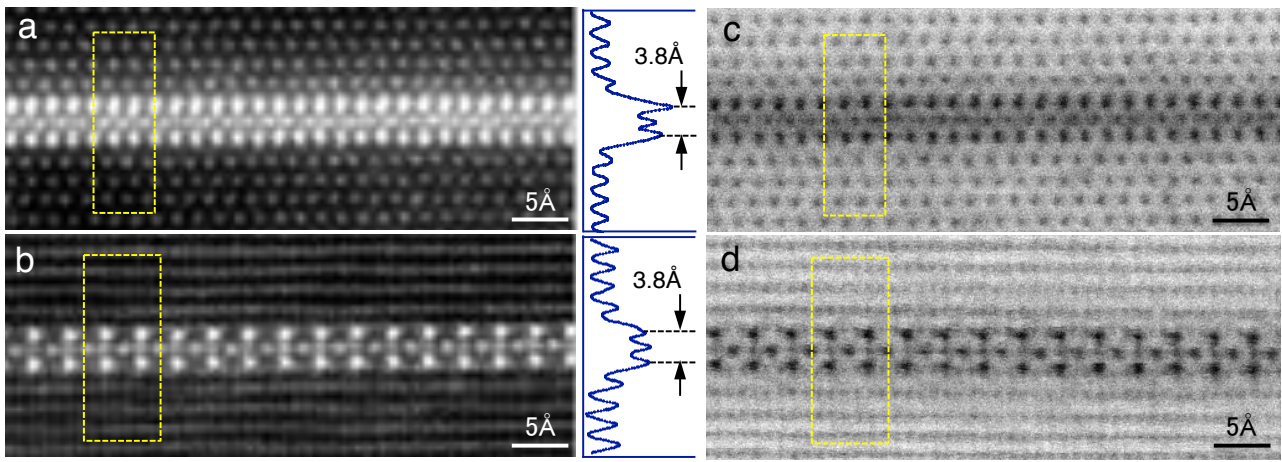

**Figure 2.** Atomic-resolution (a), (b) HAADF- and (c), (d) ABF-STEM images of the platelet-crystalline zone with a three close-packed layer thick, taken along (a), (c) $[11\bar{2}0]_{hcp}$ and (b), (d) $[1\bar{1}00]_{hcp}$ directions. Dashed rectangles represent the super-cell used for averaging procedure (Figs. 5 a, c, f, h). The layer-by-layer HAADF intensity profiles along the *c*-axis are obtained by integrating over the entire area, and the distance between the brightest layers (3.8Å) is estimated by referring the $(0002)_{Mg}$ intervals as being 2.6Å.

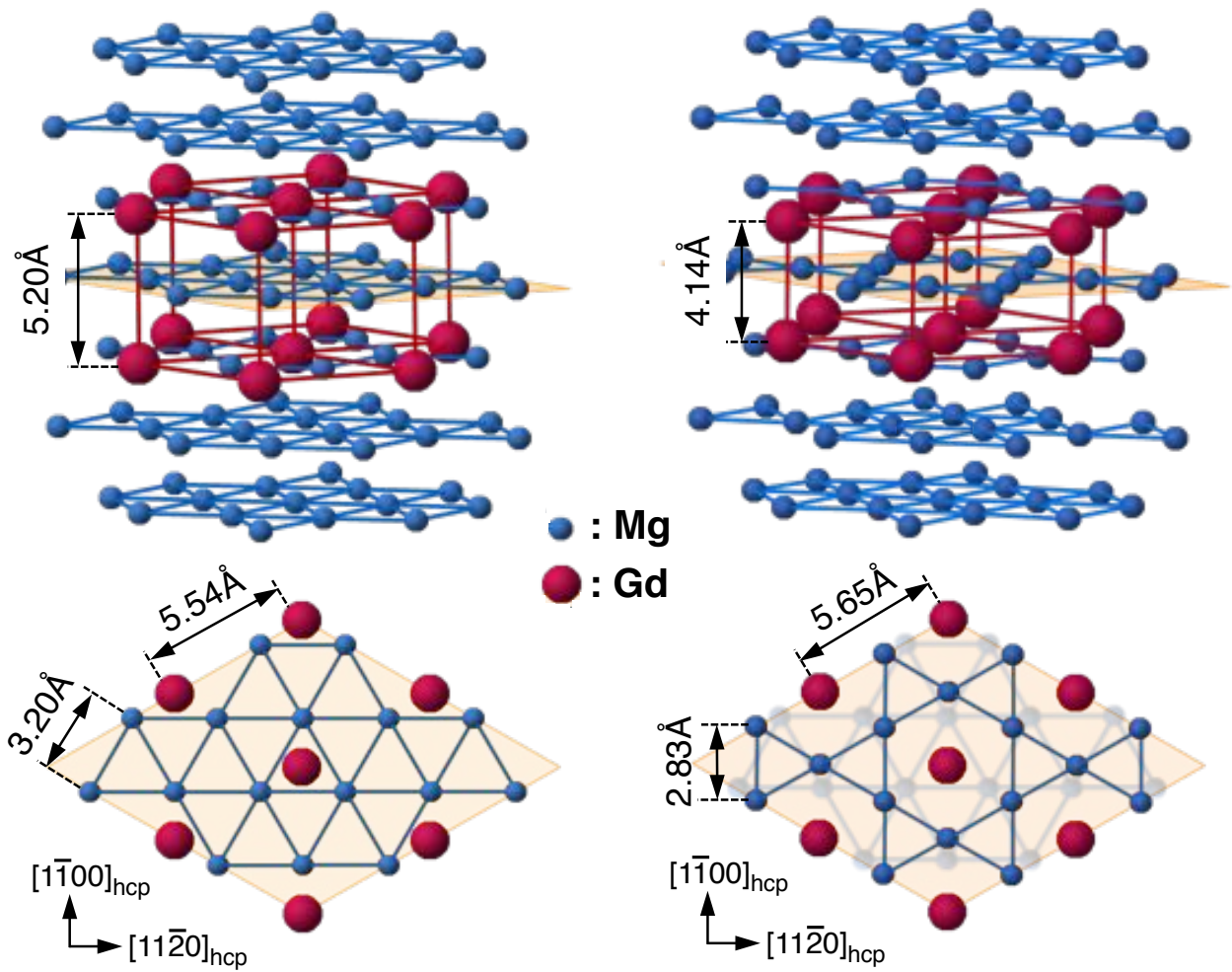

**Figure 3.** Structures of the PC zone; (left) initial configuration [16] and (right) energetically optimized configuration. First principles calculations using the VASP were made based on the $\sqrt{3}\times\sqrt{3}\times4$ super-cell of the hcp-Mg structure, with the cut-off energy 360 eV, 5×5×1 k-point mesh and Methfessel-Paxton smearing method with a width of 0.2 eV. In the bottom, Mg-atom networks at the middle close-packed layer (orange-color) are shown, together with the projected Gd atom positions; (right) the initial Mg-atom network is shown by transparent for comparison.

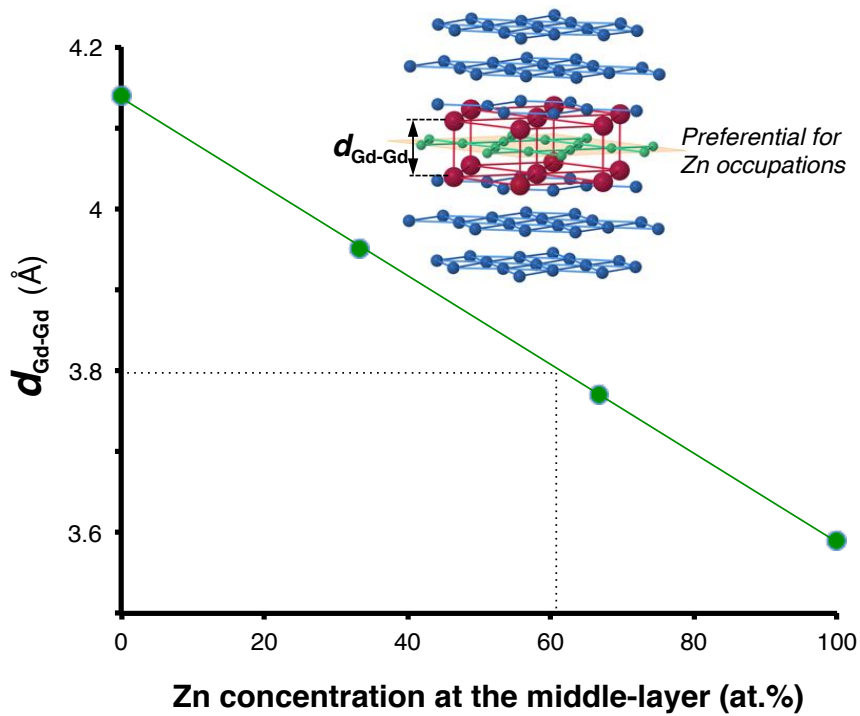

**Figure 4.** Calculations of the interatomic Gd-Gd distance along the *c*-axis, $d_{Gd\text{-}Gd}$, by changing the Zn concentrations at the middle layer of the PC zone. Calculations were made for the Zn concentrations of 0, 33.3, 66.7 and 100% at the middle layer. There is a clear linear correlation between $d_{Gd\text{-}Gd}$ and Zn concentrations, providing $d_{Gd\text{-}Gd}$ ~3.8Å when the Zn atoms occupy the middle-layer by ~60% (dashed line).

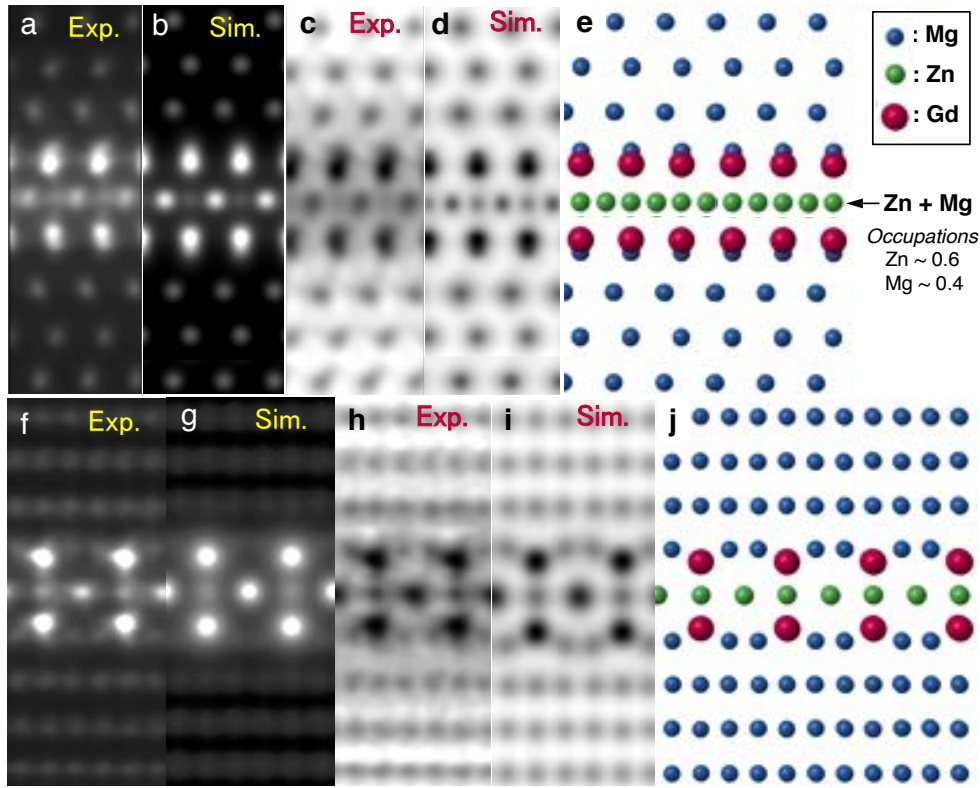

**Figure 5.** Atomic-resolution STEM images (a, c, f, h), simulated images (b, d, g, i) and the corresponding projections of the model structure (e, j). The PC zone is viewed along (a-e) $[11\bar{2}0]_{hcp}$ and (c-j) $[1\bar{1}00]_{hcp}$ directions. Experimental HAADF (a, f) and ABF (c, h) STEM images are reconstructed by averaging the super-cell regions across the entire images shown in Fig. 2. For averaging, local image distortion has been corrected using the linear-distortion matrix, and the super-cell regions have been summed up according to the cross-correlation evaluation of local contrast fluctuations (for details of the procedure, see the supplement of Ref. 19). Simulated images based on the present PC zone structure are shown for (b, g) HAADF and (d, i) ABF, calculated with the parameters; a spherical aberration coefficient $C_s = 0$ mm, specimen thickness $t \sim 15$ nm. The present structure model is shown by projections along (e) $[11\bar{2}0]_{hcp}$ and (j) $[1\bar{1}00]_{hcp}$.